\documentclass[11pt,twoside]{article}
\usepackage{asp2014}

\aspSuppressVolSlug
\resetcounters

\begin{document}

\title{Using the Astrophysics Source Code Library: Find, cite, download, parse, study, and submit}

\author{Alice~Allen}
\affil{Astrophysics Source Code Library, Houghton, MI, US; \email{aallen@ascl.net}}
\affil{University of Maryland College Park, College Park, MD, US}

\paperauthor{Alice~Allen}{aallen@ascl.net}{0000-0003-3477-2845}{}{Author1 Department}{City}{State/Province}{Postal Code}{Country}




  
\begin{abstract}

The Astrophysics Source Code Library (ASCL) contains 3000 metadata records about astrophysics research software and serves primarily as a registry of software, though it also can and does accept code deposit. Though the ASCL was started in 1999, many astronomers, especially those new to the field, are not very familiar with it. This hands-on virtual tutorial was geared to new users of the resource to teach them how to use the ASCL, with a focus on finding software and information about software not only in this resource, but also by using Google and NASA's Astrophysics Data System (ADS). With computational methods so important to research, finding these methods is useful for examining (for transparency) and possibly reusing the software (for reproducibility or to enable new research). Metadata about software is useful for, for example, knowing how to cite software when it is used for research and studying trends in the computational landscape. Though the tutorial was primarily aimed at new users, advanced users were also likely to learn something new. 

\end{abstract}

\section{Introduction and instructions for participants}
Overall, using the ASCL is straight-forward, but how it interacts with other resources, such as ADS, may not be apparent to new or casual users, and certainly there are tips and tricks for using it more efficiently and effectively. This tutorial was developed to teach new users of the ASCL how this resource works. Prospective attendees were provided with instructions, via the ADASS website, as to what they needed to have to participate in the tutorial (a computer with a browser and Internet access) and what knowledge they were assumed to already have (rudimentary familiarity with using ADS for searching for articles/resources; what a bibcode is; how to use a general search engine such as Google). They were also asked to bookmark several URLs in their browser, including the ASCL home page (\url{https://ascl.net}), the ADS home page (\url{https://ui.adsabs.harvard.edu/}, and the Google search page (\url{https://www.google.com}).

\begin{figure}
    \centering
    \includegraphics [scale=0.5]{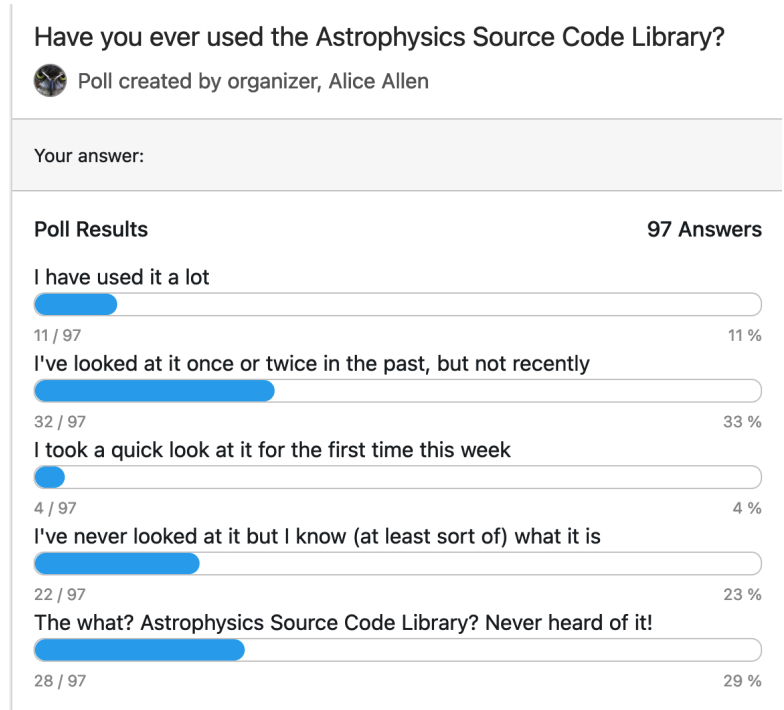}
    \caption{Poll results: Familiarity with the ASCL}
    \label{fig:ASCLfamiliarity}
\end{figure}

Upon entering the virtual space in which the tutorial was given, participants were asked to answer a poll to gauge their starting familiarity with the ASCL; Figure \ref{fig:ASCLfamiliarity} shows that most attendees had little experience with the resource.

\section{Tutorial outline}
The tutorial covered:
\begin{itemize}
    \item what the ASCL is and components of an ASCL entry
    \item common and alternate ways to bring up ASCL records
    \item how to find software using different methods and tools
    \item how citation tracking and preferred citation work
    \item how to find a code's preferred citation (where one exists)
    \item how to create a metadata file that informs others how to cite your code 
    \item the best place(s) to put preferred citation information
\end{itemize}

\section{Hands-on activities}
\begin{figure}
     \centering
     \includegraphics [scale=0.47]{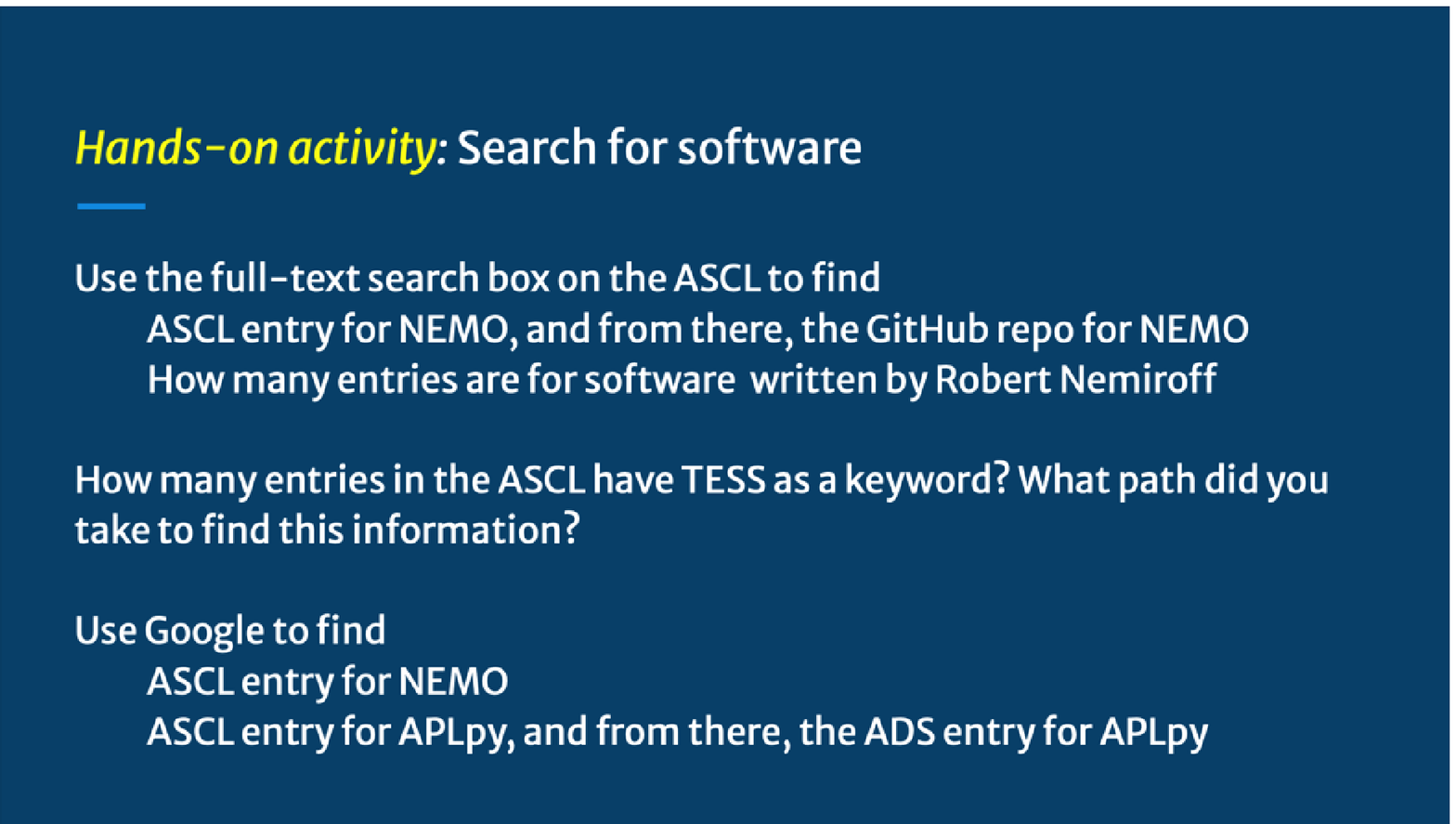}
     \caption{Hands-on Activity \#1}
     \label{fig:Searching1}
\end{figure}

\begin{figure}
     \centering
     \includegraphics [scale=0.47]{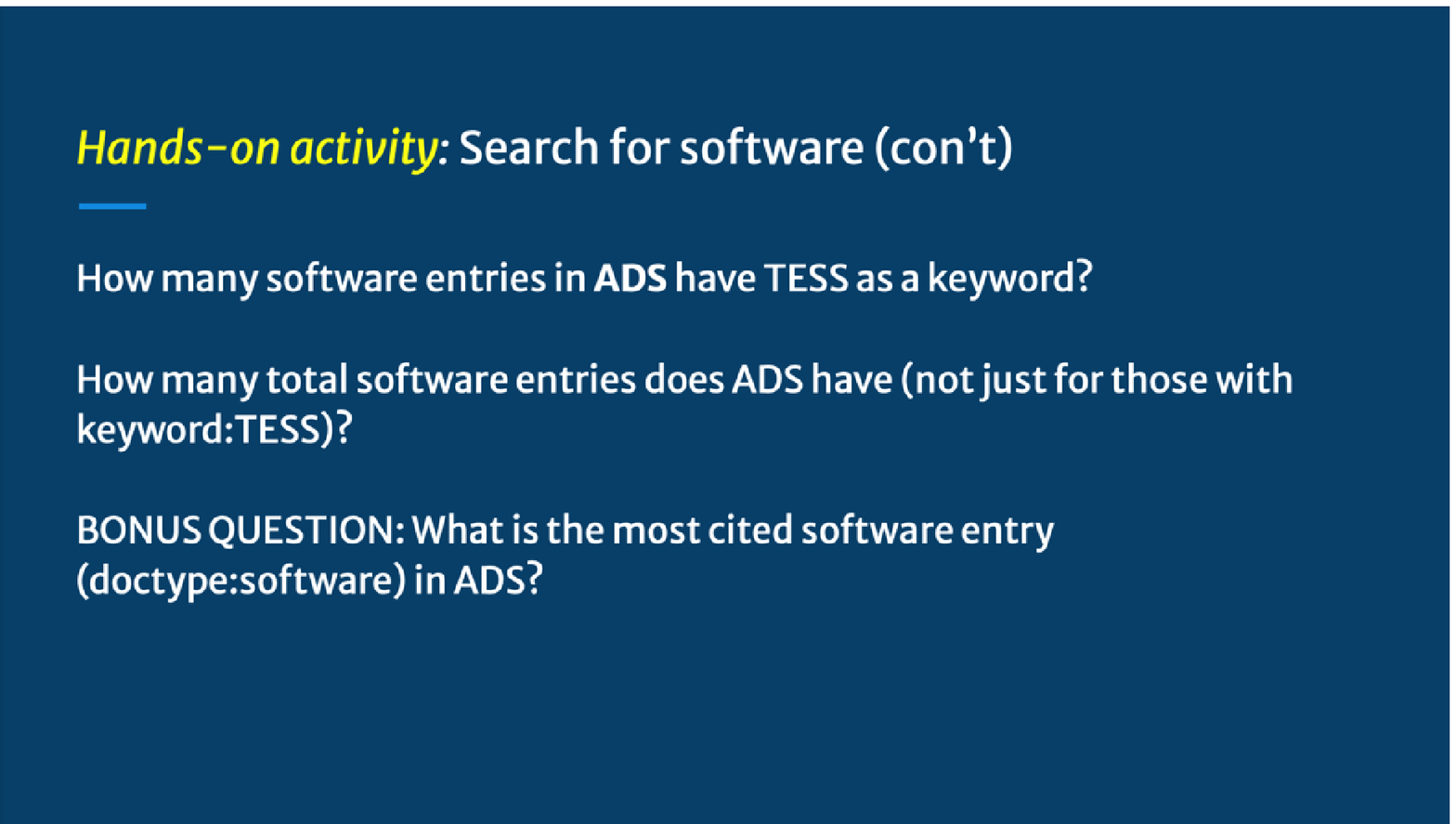}
     \caption{Hands-on Activity \#2}
     \label{fig:Searching2}
\end{figure}

Tutorial participants were encouraged to follow along and mirror the presenter's actions as different tasks, such as bringing up ASCL records or finding software in ADS, were demonstrated.  Four specific hands-on activities provided practice with the ASCL, and with using ADS and Google, too, for information relating to the ASCL, reinforcing the training. Figures \ref{fig:Searching1} and \ref{fig:Searching2} show the searching activities. Readers are encouraged to try these exercises to gauge their familiarity with the ASCL and even with ADS. The full set of slides is available on the ASCL (\url{https://tinyurl.com/ASCLtutorial}).

\section{Questions, answers, and comments}
Attendees were encouraged to post questions and comments to the Q\&A and chat areas provided by the virtual tool used to present the tutorial, and to unmute themselves to ask questions or make comments directly. Questions were varied, but three topics were of particular interest: keywords, citing software, and submitting software to ASCL. The ASCL currently has keywords only for NASA missions and HITS software. Participants asked, \textit{``Is there a guideline on list of keywords?''}, \textit{``Do you have a taxonomy for  your keywords?''}, \textit{``Can we suggest more?''}, and \textit{``Can we suggest keywords to you for our own codes?''} I was glad to learn there is so much interest in the ASCL providing more keywords! How to do that, and what keywords to use, will be discussed with the community at a later date. Regarding citation, one attendee asked, \textit{``What would you define as a good method?''}. Criteria for inclusion in the ASCL came up in questions on whether there is a way for a new code to be added to the resource before research that describes or uses the software is published (yes, there is), and what is considered a refereed resource, with the point made that SPIE proceedings, for example, may not be refereed in the traditional sense. 

In addition, one participant mentioned, \textit{``I'm really interested in these new sort options, I didn't realise ADS had added those.''} This was not surprising to me; I include information on ADS searching when I show people how to use the ASCL because so many do not know about using, for example, doctype to find just software. 

\section{Using every minute}
\begin{figure}
    \centering
    \includegraphics [scale=0.5]{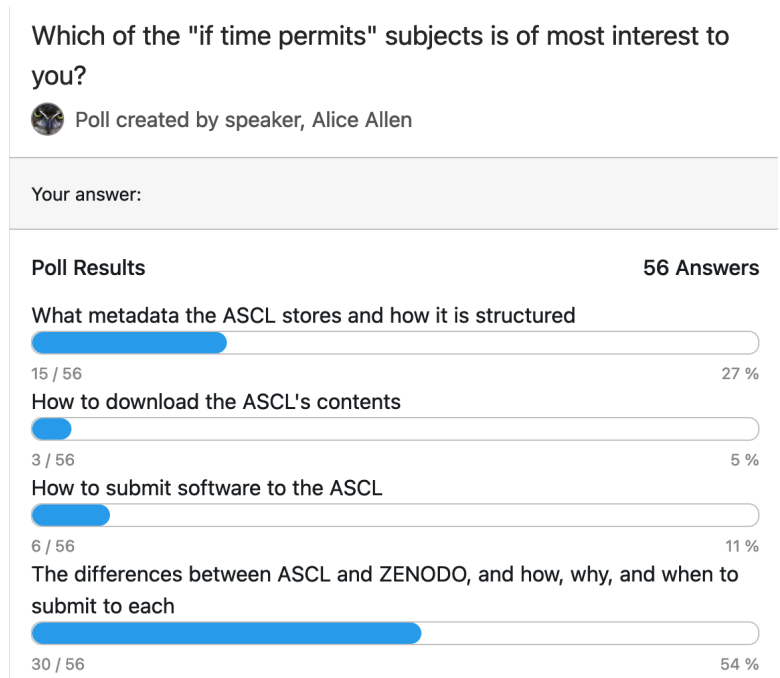}
    \caption{Poll results: Final topic for discussion}
    \label{fig:LastTopic}
\end{figure}

Four additional optional topics were made available that could be discussed if time permitted; participants were asked to express which of these was of the most interest to them in a second poll, as shown in Figure \ref{fig:LastTopic}. As a result, the differences between and similarities of ASCL and Zenodo was the final topic of the tutorial.

\section{Summary}
This tutorial was intended for people relatively new to the ASCL; polling the audience before the start of the session showed that it was on-target as to who would attend. Only 11\% of the participants had any significant experience with the resource. I thank the Heidelberg Institute for Theoretical Studies, Michigan Technological University, and the University of Maryland College Park for support, the ADASS POC for selecting this tutorial, the participants for coming, and software authors everywhere for providing the computational methods on which research depends.




\end{document}